\documentstyle[preprint,epsfig,eqsecnum,aps,floats,tighten]{revtex}

\def\Missing#1#2{{\mbox{$#1\kern-0.57em\raise0.19ex\hbox{/}_{#2}$}}}

\def\vMissing#1#2{\ifmmode
            \vec{#1}\kern-0.57em\raise.19ex\hbox{/}_{#2}
         \else
            {{\mbox{$\vec{#1}\kern-0.57em\raise.19ex\hbox{/}_{#2}$}}}
         \fi}
\def\lsim{\mathrel{\rlap{\lower4pt\hbox{\hskip1pt$\sim$}}
    \raise1pt\hbox{$<$}}}        
\def\gsim{\mathrel{\rlap{\lower4pt\hbox{\hskip1pt$\sim$}}
    \raise1pt\hbox{$>$}}}        
\def\et{\mbox{$E_{T}\,$}}

\def\pt{\mbox{$p_{T}\,$}}

\def\D0{D\O }

\newcommand{\Rsep}{\mbox{${\mathcal{R}}_{\rm sep}$}}

\def\simge{\mathrel{\rlap{\raise 0.53ex \hbox{$>$}}%
{\lower 0.53ex \hbox{$\sim$}}}}
\def\simle{\mathrel{\rlap{\raise 0.53ex \hbox{$<$}}%
{\lower 0.53ex \hbox{$\sim$}}}}

\def\ETmiss{\mbox{${\hbox{$E$\kern-0.5em\lower-.1ex\hbox{/}\kern+0.15em}}_T$ }}

\def\err#1#2#3 {{\it Erratum} {\bf#1},{\ #2} (19#3)}
\def\ib#1#2#3 {{\it ibid.} {\bf#1},{\ #2} (19#3)}
\def\nc#1#2#3 {Nuovo Cim. {\bf#1} ,#2(19#3)}
\def\nim#1#2#3 {Nucl. Instr. Meth. {\bf#1},{\ #2} (19#3)}
\def\np#1#2#3 {Nucl. Phys. {\bf#1},{\ #2} (19#3)}
\def\pl#1#2#3 {Phys. Lett. {\bf#1},{\ #2} (19#3)}
\def\prev#1#2#3 {Phys. Rev. {\bf#1},{\ #2} (19#3)}
\def\prl#1#2#3 {Phys. Rev. Lett. {\bf#1},{\ #2} (19#3)}
\def\rmp#1#2#3 {Rev. Mod. Phys. {\bf#1},{\ #2} (19#3)}
\def\zp#1#2#3 {Zeit. Phys. {\bf#1},{\ #2} (19#3)}     
\def\kt{{\it k}$_{\perp}\:$}
\begin{document}

%
%
\title{The Inclusive Jet Cross Section in 
$p\overline{p}$ Collisions at $\sqrt{s}=1.8$~TeV 
using the \kt Algorithm}
\author{\centerline{The D\O\ Collaboration
  \thanks{Submitted to the {\it International Europhysics Conference
  on High Energy Physics},
  \hfill\break
  July 12-18, 2001, Budapest, Hungary,
  \hfill\break 
  and  {\it XX International Symposium on Lepton and Photon Interactions at High Energies}
  \hfill\break
  July 23 -- 28, 2001, Rome, Italy. 
}}}
\address{
\centerline{Fermi National Accelerator Laboratory, Batavia, Illinois 60510}
}
%
%
\date{\today}

\maketitle

%
%
\begin{abstract}
We present a preliminary measurement of the central inclusive jet 
cross section using a successive combination algorithm based on 
relative transverse momenta (\kt) for jet reconstruction.
We analyze a $87.3\,pb^{-1}$ data sample
collected by the \D0 detector at the Fermilab Tevatron $p\overline{p}$
Collider during 1994-1995.  The cross section, reported as a function
of transverse momentum ($p_T > 60$ GeV) in the central region of
pseudo-rapidity ($|\eta| < 0.5$), is in reasonable agreement with
next-to-leading order QCD predictions. This is the first jet
production measurement in a hadron collider using a successive
combination type of jet algorithm.
\end{abstract}

\newpage
\begin{center}
\small{
%
V.M.~Abazov,$^{23}$                                                           
B.~Abbott,$^{58}$                                                             
A.~Abdesselam,$^{11}$                                                         
M.~Abolins,$^{51}$                                                            
V.~Abramov,$^{26}$                                                            
B.S.~Acharya,$^{17}$                                                          
D.L.~Adams,$^{60}$                                                            
M.~Adams,$^{38}$                                                              
S.N.~Ahmed,$^{21}$                                                            
G.D.~Alexeev,$^{23}$
% ...
\\                                                                            
\vskip 0.30cm                                                                 
\centerline{(D\O\ Collaboration)}                                             
\vskip 0.30cm                                                                 
\centerline{$^{1}$Universidad de Buenos Aires, Buenos Aires, Argentina}       
\centerline{$^{2}$LAFEX, Centro Brasileiro de Pesquisas F{\'\i}sicas,         
                  Rio de Janeiro, Brazil}                                     
\centerline{$^{3}$Universidade do Estado do Rio de Janeiro,                   
                  Rio de Janeiro, Brazil}                                     
\centerline{$^{4}$Institute of High Energy Physics, Beijing,                  
                  People's Republic of China}                                 
\centerline{$^{5}$Universidad de los Andes, Bogot\'{a}, Colombia}             
\centerline{$^{6}$Charles University, Center for Particle Physics,            
                  Prague, Czech Republic}                                     
\centerline{$^{7}$Institute of Physics, Academy of Sciences, Center           
                  for Particle Physics, Prague, Czech Republic}               
\centerline{$^{8}$Universidad San Francisco de Quito, Quito, Ecuador}         
\centerline{$^{9}$Institut des Sciences Nucl\'eaires, IN2P3-CNRS,             
                  Universite de Grenoble 1, Grenoble, France}                 
\centerline{$^{10}$CPPM, IN2P3-CNRS, Universit\'e de la M\'editerran\'ee,     
                  Marseille, France}                                          
\centerline{$^{11}$Laboratoire de l'Acc\'el\'erateur Lin\'eaire,              
                  IN2P3-CNRS, Orsay, France}                                  
\centerline{$^{12}$LPNHE, Universit\'es Paris VI and VII, IN2P3-CNRS,         
                  Paris, France}                                              
\centerline{$^{13}$DAPNIA/Service de Physique des Particules, CEA, Saclay,    
                  France}                                                     
\centerline{$^{14}$Universit{\"a}t Mainz, Institut f{\"u}r Physik,            
                  Mainz, Germany}                                             
\centerline{$^{15}$Panjab University, Chandigarh, India}                      
\centerline{$^{16}$Delhi University, Delhi, India}                            
\centerline{$^{17}$Tata Institute of Fundamental Research, Mumbai, India}     
\centerline{$^{18}$Seoul National University, Seoul, Korea}                   
\centerline{$^{19}$CINVESTAV, Mexico City, Mexico}                            
\centerline{$^{20}$FOM-Institute NIKHEF and University of                     
                  Amsterdam/NIKHEF, Amsterdam, The Netherlands}               
\centerline{$^{21}$University of Nijmegen/NIKHEF, Nijmegen, The               
                  Netherlands}                                                
\centerline{$^{22}$Institute of Nuclear Physics, Krak\'ow, Poland}            
\centerline{$^{23}$Joint Institute for Nuclear Research, Dubna, Russia}       
\centerline{$^{24}$Institute for Theoretical and Experimental Physics,        
                   Moscow, Russia}                                            
\centerline{$^{25}$Moscow State University, Moscow, Russia}                   
\centerline{$^{26}$Institute for High Energy Physics, Protvino, Russia}       
\centerline{$^{27}$Lancaster University, Lancaster, United Kingdom}           
\centerline{$^{28}$Imperial College, London, United Kingdom}                  
\centerline{$^{29}$University of Arizona, Tucson, Arizona 85721}              
\centerline{$^{30}$Lawrence Berkeley National Laboratory and University of    
                  California, Berkeley, California 94720}                     
\centerline{$^{31}$University of California, Davis, California 95616}         
\centerline{$^{32}$California State University, Fresno, California 93740}     
\centerline{$^{33}$University of California, Irvine, California 92697}        
\centerline{$^{34}$University of California, Riverside, California 92521}     
\centerline{$^{35}$Florida State University, Tallahassee, Florida 32306}      
\centerline{$^{36}$University of Hawaii, Honolulu, Hawaii 96822}              
\centerline{$^{37}$Fermi National Accelerator Laboratory, Batavia,            
                   Illinois 60510}                                            
\centerline{$^{38}$University of Illinois at Chicago, Chicago,                
                   Illinois 60607}                                            
\centerline{$^{39}$Northern Illinois University, DeKalb, Illinois 60115}      
\centerline{$^{40}$Northwestern University, Evanston, Illinois 60208}         
\centerline{$^{41}$Indiana University, Bloomington, Indiana 47405}            
\centerline{$^{42}$University of Notre Dame, Notre Dame, Indiana 46556}       
\centerline{$^{43}$Iowa State University, Ames, Iowa 50011}                   
\centerline{$^{44}$University of Kansas, Lawrence, Kansas 66045}              
\centerline{$^{45}$Kansas State University, Manhattan, Kansas 66506}          
\centerline{$^{46}$Louisiana Tech University, Ruston, Louisiana 71272}        
\centerline{$^{47}$University of Maryland, College Park, Maryland 20742}      
\centerline{$^{48}$Boston University, Boston, Massachusetts 02215}            
\centerline{$^{49}$Northeastern University, Boston, Massachusetts 02115}      
\centerline{$^{50}$University of Michigan, Ann Arbor, Michigan 48109}         
\centerline{$^{51}$Michigan State University, East Lansing, Michigan 48824}   
\centerline{$^{52}$University of Nebraska, Lincoln, Nebraska 68588}           
\centerline{$^{53}$Columbia University, New York, New York 10027}             
\centerline{$^{54}$University of Rochester, Rochester, New York 14627}        
\centerline{$^{55}$State University of New York, Stony Brook,                 
                   New York 11794}                                            
\centerline{$^{56}$Brookhaven National Laboratory, Upton, New York 11973}     
\centerline{$^{57}$Langston University, Langston, Oklahoma 73050}             
\centerline{$^{58}$University of Oklahoma, Norman, Oklahoma 73019}            
\centerline{$^{59}$Brown University, Providence, Rhode Island 02912}          
\centerline{$^{60}$University of Texas, Arlington, Texas 76019}               
\centerline{$^{61}$Texas A\&M University, College Station, Texas 77843}       
\centerline{$^{62}$Rice University, Houston, Texas 77005}                     
\centerline{$^{63}$University of Virginia, Charlottesville, Virginia 22901}   
\centerline{$^{64}$University of Washington, Seattle, Washington 98195}       

%
\bibitem[*]{lehner}
Visitor from University of Zurich, Zurich, Switzerland.
\vskip 0.25cm

}

\end{center}

\normalsize

\vfill\eject

\section{Introduction}
\label{sec:intro}
Jet production in hadronic collisions is understood within the
framework of Quantum Chromodynamics (QCD) as a hard scattering of 
constituent partons (quarks and gluons), which, having undergone the
interaction, manifest themselves as showers of collimated
particles called jets. 
Jet algorithms associate clusters of these particles into
jets in a way that the kinematic properties of the hard-scattered
partons can be inferred and thereby compared to predictions from
perturbative QCD (pQCD).


Historically, only the cone algorithm has been used 
to reconstruct jets at hadron colliders~\cite{snowmass}.  Although
well suited to implement the experimental corrections needed in the
complex environment of hadron colliders, the cone algorithms used
in previous measurements by the hadron collider experiments at the
Fermilab Tevatron present several difficulties, because (a) an
arbitrary procedure must be implemented to split and merge overlapping
cones, (b) an ad-hoc parameter, \Rsep, is required to accommodate the
differences between jet definitions at the parton and detector
levels~\cite{rsep}, and (c) improved theoretical predictions
calculated at the Next-to-Next-to-Leading-Order (NNLO) in pQCD are not
infrared safe, because they exhibit a marked sensitivity to soft
radiation~\cite{soft_rad}.

Inspired by QCD, a second class of jet algorithms, which does not
suffer from these shortcomings, has been developed by several
groups~\cite{catani93,catani92,ellis93}.  These
clustering or recombination algorithms successively merge pairs of
nearby vectors (partons, particles or calorimeter towers) in order of
increasing relative transverse momentum ({\it p}$_{\mathrm{T}}$).  A
single arbitrary parameter, $D$, which characterizes approximately the
size of the resulting jets, is used to determine when this merging
stops.  No splitting-merging is involved, because each vector is
assigned to a unique jet. There is no need for introducing any ad-hoc
parameters, because the same algorithm is applied at the theoretical
and experimental level.  Furthermore, by design, clustering algorithms
are infrared and collinear safe to all orders of calculation.
The dependence of the inclusive jet cross section on the choice of
reconstruction algorithms or parameters is particularly relevant for
studying the effect of hadronization and background from spectator
partons in the event.
In contrast to previous work from hadron
colliders~\cite{cdf,d0cone,d0levan,d0_jets_prd,cdf_jets_prd}, this
paper presents the first measurement of the inclusive jet cross
section using the \kt algorithm to reconstruct jets.

\section{Jet Reconstruction and Selection}
\label{sec:xsec}

The differential jet cross section was measured in bins of \pt and
pseudo-rapidity, $\eta \equiv -{\rm ln}[{\rm tan}(\theta/2)]$, where a
right handed coordinate system is adopted with the $z$ axis pointing
in the proton beam direction, and $\theta$ is the polar angle.  The
\kt algorithm implemented at D\O ~\cite{rob_prd} is based on the
clustering algorithm suggested in Ref.~\cite{ellis93}.  The algorithm
starts with a list of pre-clusters or ``vectors''. For each vector
$p_{T,i}$, $d_{ii}=p_{T,i}^2$ and for each pair of vectors,
$d_{ij}=min(p_{T,i}^2,p_{T,j}^2)\,\Delta R_{i,j}^2/D^2$ are defined,
where D is the free parameter of the algorithm and $\Delta
R_{i,j}^2=\Delta\phi_{ij}^2+\Delta\eta_{ij}^2$ is the square of the
angular separation between the vectors.
If the minimum of all $d_{ii}$ and $d_{ij}$ is a $d_{ij}$, then the
vectors $i$ and $j$ are merged, becoming the merged four-vector
$(E_i+E_j,\,\vec{p}_i+\vec{p}_j)$.  If the minimum is a $d_{ii}$, the
vector $i$ is defined as a jet.  This procedure is repeated until all
vectors are combined into jets.  Thus \kt jets do not have to include
all vectors in a cone of radius $D$, and can include vectors outside
of this cone.

The primary tool for jet detection at D\O\ is the uranium-liquid argon
calorimeter~\cite{d0calor}, which has full coverage for
pseudo-rapidity $|\eta|<4.1$.
The initial hardware trigger selected inelastic collisions as defined
by hodoscopes located near the beam axis on both sides of the
interaction region. The next stage required energy deposition in any
$\Delta\eta\times\Delta\phi=0.8\times 1.6$ region of the calorimeter,
corresponding to a transverse energy ({\it E}$_{\mathrm{T}}$) above a
preset threshold. Selected events were digitized and sent to an array
of processors. Jet candidates were reconstructed with a cone algorithm
(with radius $R=0.7$), and the event was recorded if any jet
\et exceeded a specified threshold. Jet \et thresholds of $30$, $50$, 
$85$ and $115$ GeV were used to accumulate integrated luminosities of
$0.34$, $4.46$, $51.5$ and $87.3$ pb$^{-1}$, 
respectively~\cite{d0_jets_prd}.

Jets were reconstructed offline using the \kt algorithm, with
$D=1.0$. This value of $D$ was chosen because, at 
next-to-leading-order (NLO), it produces a
theoretical cross section that is essentially identical to the cone
prediction for $R=0.7$~\cite{ellis93}, as used by 
D\O\ in its previous publications on jet production~\cite{d0_jets_prd}.
The imbalance in transverse momentum, ``missing transverse energy'', 
was calculated from the vector sum of
the \et values in all cells of the calorimeter.
The vertices of the events were reconstructed using the central tracking
system~\cite{d0calor}. A significant portion of the data was taken at high
instantaneous luminosity, where more than one interaction per beam
crossing was probable. 
When an event had more than
one reconstructed vertex, the quantity $S_T=|\Sigma \vec{p}_T^{\,jet}|$ was
defined for the two vertices that had the largest number of tracks, and
the vertex with the smaller $S_T$ was retained 
for calculating all kinematic variables.
To preserve the
pseudo-projective nature of the D\O\ calorimeter, the selected vertex
was required to be within $50$ cm of the center of the detector. This
requirement rejected $(10.6\pm0.1)\%$ of the events, independent of jet
transverse momentum.

Isolated noisy calorimeter cells were suppressed using online and
offline algorithms~\cite{thesis}.  Background introduced by electrons,
photons, detector noise and accelerator losses that mimicked jets were
eliminated with quality cuts. The efficiency of jet selection was
approximately $99.5\%$ and nearly independent of jet \pt.
Background events from cosmic rays or misvertexed events were
eliminated by requiring the missing transverse energy in each event to
be less than $70\%$ of the \pt of the leading jet.  This criterion was
nearly $100\%$ efficient.


The D\O\ jet momentum calibration~\cite{rob_prd}, applied on a jet by
jet basis, corrects on average the reconstructed \pt for
background from spectator partons (the ``underlying event'',
determined from minimum-bias events), additional interactions, pileup
from previous $p\overline{p}$ crossings, noise from uranium
radioactivity, detector non-uniformities, and for the global response
of the detector to hadronic jets.  Unlike the cone algorithm, the \kt
algorithm does not require additional corrections for showering in the
calorimeter~\cite{rob_prd}. For $|\eta|<0.5$, the mean total
multiplicative correction factor to an observed \pt of $100$ GeV
[$400$ GeV] was $1.094\pm0.015$ [$1.067\pm0.020$].

\section{Inclusive Cross Section}
\label{sec:results}

The inclusive jet cross section for $|\eta|<0.5$ was calculated
in four ranges of transverse momentum, using data from only one 
trigger in each case. The more restrictive trigger was used as
soon as it became fully efficient.
The average differential cross section for each \pt bin,
$d^2\sigma/(dp_T d\eta)$, was calculated as $N/(\Delta\eta \Delta p_T
\epsilon \int L dt)$, where $\Delta\eta$, $\Delta p_T$ are the $\eta$,
$p_T$ bin sizes, $N$ the number of jets observed in that bin,
$\epsilon$ is the overall efficiency for jet and event selection, and
$\int L dt$ represents the integrated luminosity of the data sample.

The measured cross section is distorted in \pt by the momentum
resolution of the D\O\ calorimeter. Although the resolution in jet \pt
is essentially Gaussian, the steepness of the \pt spectrum shifts the
observed cross section to larger values.  The fractional momentum
resolution was determined from the imbalance in \pt in two-jet
events~\cite{thesis}.  At $100$ GeV [$400$ GeV] the fractional
resolution was $0.061\pm0.006$ [$0.039\pm0.003$].  The distortion in
the cross section due to the resolution was corrected by assuming an
ansatz function, $A\,p_T^{-B} (1-2\,p_T/\sqrt{s}\,)^C$, smearing this
with the measured resolution, and fitting the parameters $A$, $B$ and
$C$ so as to best describe the observed cross section.
The bin-to-bin ratio of the original
ansatz to the smeared one was used to remove the distortion due to
resolution. The unsmearing correction reduces the
observed cross section by $(5.7\pm1)\%$ [$(6.1\pm1)\%$] at $100$ GeV
[$400$ GeV].

\begin{figure}[htbp]
\centerline{\psfig{figure=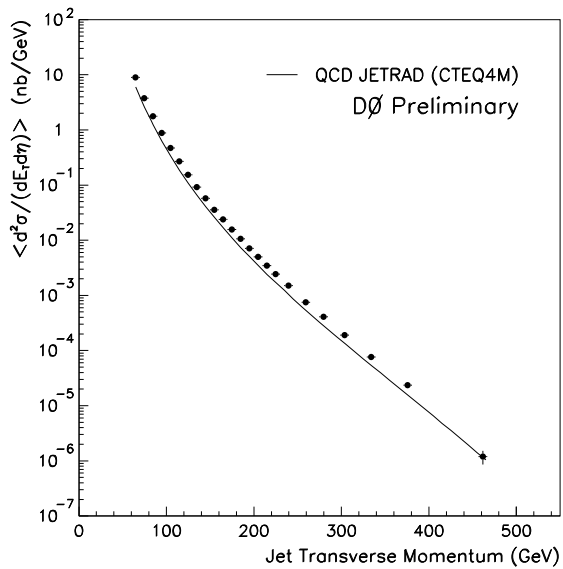,width=3.25in}}
\caption{The central ($|\eta|<0.5$) inclusive jet cross 
section obtained with the \kt algorithm at $\sqrt{s}=1.8$ TeV. 
Only statistical errors are included. The solid line shows a 
prediction from NLO pQCD.}
\label{fig:xsec}
\end{figure}

\begin{figure}[htbp]
\centerline{\psfig{figure=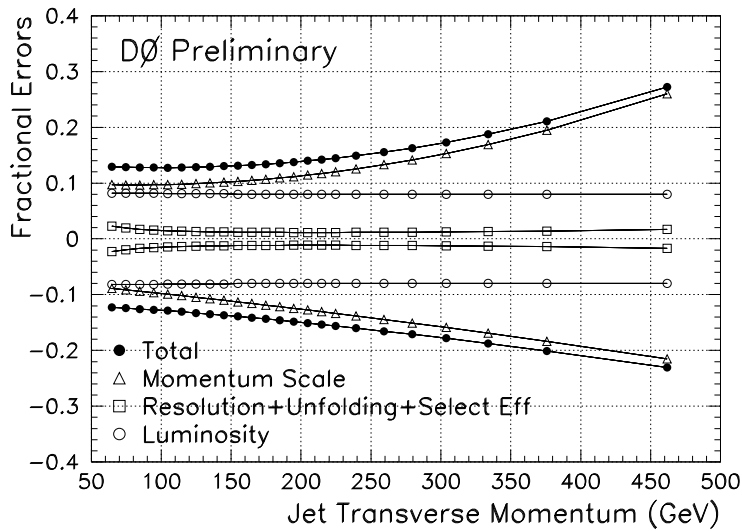,width=4.25in}}
\caption{Fractional experimental uncertainties on the cross
section.}
\label{fig:fractional_errors}
\end{figure}

The final, fully corrected, cross section for $|\eta|<0.5$ is shown in
Fig.~\ref{fig:xsec}, along with statistical uncertainties.  Listed in
Table~\ref{table:xsec} are the
\pt ranges, the point positions, the cross section, and uncertainties
in each bin.  The systematic uncertainties include contributions from
jet and event selection, unsmearing, luminosity and the uncertainty in
the momentum scale, which dominates at all transverse momenta.  The
fractional uncertainties for the different components are plotted in
Fig.~\ref{fig:fractional_errors} as a function of the jet transverse
momentum.

\section{Comparison with Theory}
\label{sec:theory}

\begin{figure}[htbp]
\centerline{\psfig{figure=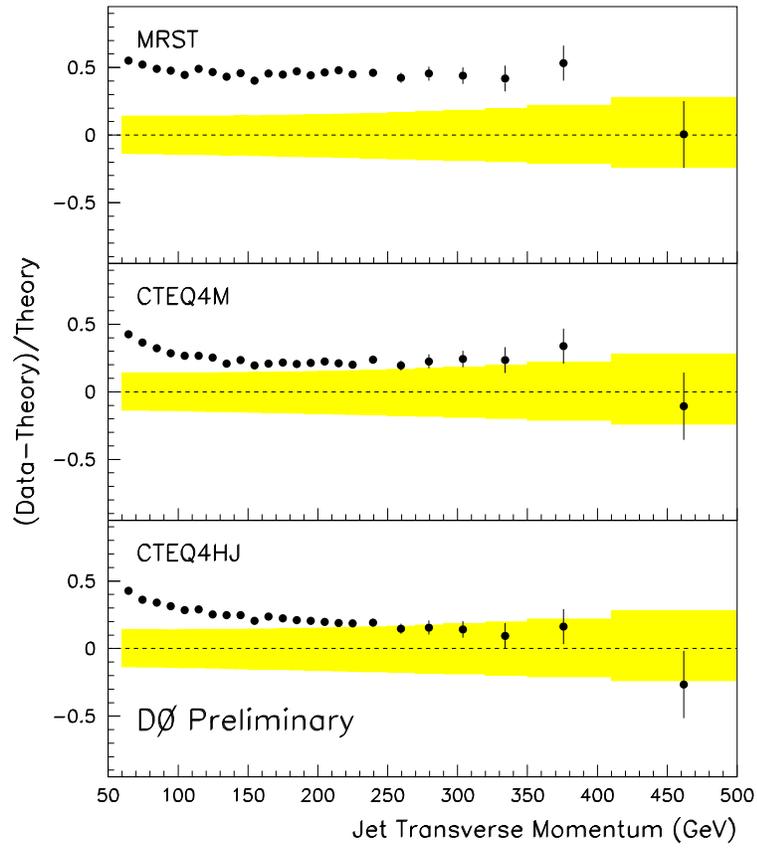,width=4.25in}}
\caption{Difference between data and JETRAD pQCD normalized to
predictions. The outer error bars represent the total systematic
uncertainty.  }
\label{fig:dtt}
\end{figure}

The results are compared to the pQCD NLO prediction from
JETRAD~\cite{jetrad}, with the renormalization and factorization
scales set to $p_T^{max}/2$, where $p_T^{max}$ refers to the \pt of
the leading jet in an event.
The comparisons are made using parametrizations of the parton
distributions functions (PDFs) of the CTEQ~\cite{cteq} and
MRST~\cite{mrs} families.  Figure~\ref{fig:dtt} shows the ratios of
$(D-T)/T$, where $D$ refers to data and $T$ to the theoretical
prediction.
To quantify the comparison in Fig.~\ref{fig:dtt},
the fractional systematic uncertainties are multiplied by the
predicted cross section and a $\chi^2$ comparison is carried out. The
results are shown in Table~\ref{table:chi2}.
The agreement is reasonable ($\chi^2/dof$ ranges from $1.56$ to
$1.12$, the probabilities from $4$ to $31\%$), although the
differences in normalization and shape, especially at low \pt, are
quite large.  The points at low \pt have the highest impact on the
$\chi^2$.  If the first four data points are not used in the $\chi^2$
comparison, the probabilities increase to the $60-80\%$ range.



%
%

\section{Conclusions}
\label{sec:conclusion}

In conclusion, 
a preliminary measurement in proton-antiproton collisions at
$\sqrt{s}=1.8$ TeV of the inclusive jet cross section based on the
\kt algorithm has been presented. The quantitative test shows reasonable 
agreement between data and NLO pQCD predictions.






%
%

\begin{table}[htbp]
\squeezetable
\begin{center}
\begin{tabular}{cccc}
\bf Bin Range &\bf Plotted &\bf Cross Sec. $\pm$ Stat. 
&\bf Systematic \\
\bf  (GeV) & \bf \pt (GeV) & \bf (nb/GeV) &\bf Uncer ($\%$) \\
\hline\hline
%
%
 $ 60 -  70$ & $ 64.6$ & $ (8.94\pm 0.06)\times 10^{0} $ & -13, +14 \\
 $ 70 -  80$ & $ 74.6$ & $ (3.78\pm 0.04)\times 10^{0} $ & -13, +14 \\
 $ 80 -  90$ & $ 84.7$ & $ (1.77\pm 0.02)\times 10^{0} $ & -13, +14 \\
 $ 90 - 100$ & $ 94.7$ & $ (8.86\pm 0.25)\times 10^{-1}$ & -13, +14 \\
 $100 - 110$ & $104.7$ & $ (4.68\pm 0.04)\times 10^{-1}$ & -14, +14 \\
 $110 - 120$ & $114.7$ & $ (2.68\pm 0.03)\times 10^{-1}$ & -14, +14 \\
 $120 - 130$ & $124.8$ & $ (1.53\pm 0.02)\times 10^{-1}$ & -14, +14 \\
 $130 - 140$ & $134.8$ & $ (9.19\pm 0.16)\times 10^{-2}$ & -14, +14 \\
 $140 - 150$ & $144.8$ & $ (5.77\pm 0.12)\times 10^{-2}$ & -14, +14 \\
 $150 - 160$ & $154.8$ & $ (3.57\pm 0.03)\times 10^{-2}$ & -15, +14 \\
 $160 - 170$ & $164.8$ & $ (2.39\pm 0.02)\times 10^{-2}$ & -15, +14 \\
 $170 - 180$ & $174.8$ & $ (1.56\pm 0.02)\times 10^{-2}$ & -15, +14 \\
 $180 - 190$ & $184.8$ & $ (1.05\pm 0.02)\times 10^{-2}$ & -15, +14 \\
 $190 - 200$ & $194.8$ & $ (7.14\pm 0.13)\times 10^{-3}$ & -16, +15 \\
 $200 - 210$ & $204.8$ & $ (4.99\pm 0.08)\times 10^{-3}$ & -16, +15 \\
 $210 - 220$ & $214.8$ & $ (3.45\pm 0.07)\times 10^{-3}$ & -16, +15 \\
 $220 - 230$ & $224.8$ & $ (2.43\pm 0.06)\times 10^{-3}$ & -16, +15 \\
 $230 - 250$ & $239.4$ & $ (1.50\pm 0.03)\times 10^{-3}$ & -17, +16 \\
 $250 - 270$ & $259.4$ & $ (7.52\pm 0.23)\times 10^{-4}$ & -17, +16 \\
 $270 - 290$ & $279.5$ & $ (4.07\pm 0.17)\times 10^{-4}$ & -18, +17 \\
 $290 - 320$ & $303.8$ & $ (1.93\pm 0.09)\times 10^{-4}$ & -18, +18 \\
 $320 - 350$ & $333.9$ & $ (7.61\pm 0.59)\times 10^{-5}$ & -19, +19 \\
 $350 - 410$ & $375.8$ & $ (2.36\pm 0.23)\times 10^{-5}$ & -20, +21 \\ 
 $410 - 560$ & $461.8$ & $ (1.18\pm 0.33)\times 10^{-6}$ & -23, +27 \\
\end{tabular}
\end{center}
\vspace*{-.2cm}
\caption{Single inclusive cross section with jets reconstructed using the
\kt algorithm in the central pseudo-rapidity region.}
\label{table:xsec}
\end{table}

\begin{table}[htbp]
\begin{center}
{\footnotesize
\begin{tabular}{l c c c}
\bf PDF & \bf \raisebox{-.0cm}{$\chi^2$} 
        & \bf \raisebox{-.0cm}{$\chi^2/dof$}  
        & \bf Probability ($\%$) \\
\hline
MRST              & 26.8 & 1.12 & 31 \\
MRSTg$\uparrow$   & 33.1 & 1.38 & 10 \\
MRSTg$\downarrow$ & 28.2 & 1.17 & 25 \\
CTEQ3M            & 37.5 & 1.56 & 4 \\
CTEQ4M            & 31.2 & 1.30 & 15 \\
CTEQ4HJ           & 27.2 & 1.13 & 29 \\
\end{tabular}
}
\end{center}
\vspace*{-.4cm}
\caption{
$\chi^2$ comparisons ($24$ degrees of freedom) between JETRAD, with
renormalization and factorization scales set to $p_T^{max}/2$, and data
for various PDFs. 
}
\label{table:chi2}
\end{table}

\section*{Acknowledgements}
\label{sec:ack}
%
%
We thank the staffs at Fermilab and collaborating institutions, 
and acknowledge support from the 
Department of Energy and National Science Foundation (USA),  
Commissariat  \` a L'Energie Atomique and 
CNRS/Institut National de Physique Nucl\'eaire et 
de Physique des Particules (France), 
Ministry for Science and Technology and Ministry for Atomic 
   Energy (Russia),
CAPES and CNPq (Brazil),
Departments of Atomic Energy and Science and Education (India),
Colciencias (Colombia),
CONACyT (Mexico),
Ministry of Education and KOSEF (Korea),
CONICET and UBACyT (Argentina),
The Foundation for Fundamental Research on Matter (The Netherlands),
PPARC (United Kingdom),
Ministry of Education (Czech Republic),
and the A.P.~Sloan Foundation.
%

\end{document}